\documentclass[fleqn,usenatbib]{mnras}

\usepackage{graphicx} 
\usepackage{amsmath} 
\usepackage[T1]{fontenc}
\usepackage{newtxtext,newtxmath}

\DeclareRobustCommand{\VAN}[3]{#2}
\let\VANthebibliography\thebibliography
\def\thebibliography{\DeclareRobustCommand{\VAN}[3]{##3}\VANthebibliography}

\title[Vortex Avalanche Pulsar Glitches]{A Vortex Avalanche Model for Pulsar Glitches}
\author[Alpar \& Gügercinoğlu]{
M. Ali Alpar$^{1}$ and Erbil Gügercinoğlu$^{2}$\thanks{E-mail: egugercinoglu@gmail.com}
\\
$^{1}$Emeritus, Faculty of Engineering and Natural Sciences, Sabancı University, Orhanlı, Tuzla, 34956 Istanbul, Türkiye\\
$^{2}$School of Arts and Sciences, Qingdao Binhai University, Huangdao District, 266555, Qingdao, People’s Republic of China
}
\date{Accepted XXX. Received YYY; in original form ZZZ}

\begin{document}
\label{firstpage}
\pagerange{\pageref{firstpage}--\pageref{lastpage}}
\maketitle
\begin{abstract} 
Pulsar glitches are thought to involve sudden unpinning and fast transport of large numbers of pinned vortices, possibly triggered by breaking of the neutron star crust solid. Such 
a vortex avalanche can take place on a fast timescale only if there are pathways for fast vortex motion. Formation of local vortex ``traps'' (regions containing larger than average 
density of pinned vortices) will provide vortex-free regions surrounding the traps, which can connect to form pathways for fast vortex motion. We propose a simple two-dimensional 
model for the neutron star crust superfluid, with the vortex pinning landscape tiled into traps and surrounding vortex-free regions which can connect to azimuthal pathways for 
fast vortex motion. The trap and tile sizes are estimated from the physics of the crust solid and vortex pinning, namely the critical strain angle of the crust solid and the 
critical velocity for vortex unpinning. Unpinned vortices move with the superfluid flow in the azimuthal direction while also scattering radially outward, by a small angle, due 
to mutual friction. The model qualitatively explains the glitch parameters and their evolution in young pulsars and provides an explanation for the bi-modality in the distribution of 
glitch magnitudes. Laboratory experiments and simulations on vortex systems with spatially inhomogeneous vortex pinning landscapes are likely to be more realistic models for pulsar 
glitches. Breaking the spatial scale-invariance and monitoring the results  may also lead to better understanding of collective behaviour in pinned vortex systems in general.

\end{abstract}

% --- From Page 2.jpeg ---


\section*{1. INTRODUCTION}

Sudden unpinning of quantized vortices has been proposed as an explanation of pulsar glitches \citep{anderson75,alpar84a}. While this model has been successful in evaluation and interpretation of glitches and inter-glitch timing behaviour \citep{alpar84a,alpar89,erbil20,erbil22}, a detailed understanding of how sudden vortex unpinning develops into an avalanche involving a large number of vortices has not been modelled \citep{haskell15,zhou22,antonopoulou22,antonelli22}. 
Laboratory experiments in terrestrial superfluids and Bose-Einstein condensates and computational models aiming to simulate pulsar glitches are also of intrinsic interest for 
understanding the collective behaviour of pinned vortex distributions in these systems. In pioneering experiments \citep{tsakadze80}, a container filled with superfluid $^{4}$He and with 
roughened inner surfaces of the container was set into rotation, to provide pinning centres inhibiting smooth motion of quantised vortices in the superfluid. The system was 
subsequently spun down by an external torque applied to the container, and the rotation and spin-down rates of the container were then monitored. Analogous experiments have been performed on vortex systems such as superfluid $^4$He and $^3$He, Bose-Einstein condensates, type II superconductors and supersolids \citep{vaneysden14,Fieldetal95,Klausetal22,poli23,marmorini24,NovotnyTalirVarga26,AuttiGraberHaskell26}. These experiments and simulations 
involve uniform distributions of vortices and pinning centres. Discrete events of spontaneous spin-up are resolved in the rotation rate of the container, which then relax 
back to constant spin-down rates. These events rise and relax on similar timescales. The recorded distributions of fractional event sizes $\Delta\Omega_{\rm c}/\Omega_{\rm c}$, where 
$\Omega_{\rm c}$ is the container rotation rate, are scale-invariant power-laws. 

However, the fractional sizes of observed pulsar glitches do not have scale-invariant distributions \citep{melatos18,fuentes19}. The distibution of 
$\Delta\Omega_{\rm c}/\Omega_{\rm c}$, where $\Omega_{\rm c}$ is the observed crust rotation rate of the neutron star, and $\Delta\Omega_{\rm c}$ the step increase of 
$\Omega_{\rm c}$ in a glitch, is bimodal for some pulsars with many observed glitches. The distribution of glitch sizes from the entire pulsar sample is also bimodal 
\citep{celora20,basu22,ananth25}. Pulsar glitches break scale-invariance also dynamically: their rise is very rapid, faster than the resolution of pulse timing, while post-glitch 
relaxation proceeds in a variety of modes and timescales, extending over years, all the way to the arrival of the next glitch \citep{espinoza11,erbil20,lower21}. What is the reason 
for the breaking of scale-invariance in pulsar glitches? 

The distribution and long-range interactions of the vortices determine the superfluid rotation rate, while the coupling between the superfluid and normal matter components of the 
star is mediated through the interactions of vortex cores with normal matter, in particular pinning of vortex lines to sites in the neutron star's solid crust and to quantized flux 
lines in the core of the neutron star where superfluid neutrons co-exist with superconducting protons \citep{alpar84a,erbil14}. In addition, there are dissipative interactions that 
depend on the relative velocity of the vortex lines with respect to normal matter$-$electrons and phonons \citep{bildsten89,jones90}. Simulations involving some $10^3 - 10^5$ 
vortices are currently emerging \citep{warszawski13,liu25,ananth26}. In between the sudden unpinning events there is a continuous transport of vortices by unpinning and re-pinning 
processes enabled by thermal activation. This ``vortex creep'' mediates the angular momentum transfer from the superfluid to normal matter, allowing the spin-down of the superfluid 
in response to the spin-down of the normal matter crust of the neutron star under the pulsar torque \citep{alpar84a,alpar89}. The vortex creep process can be used to model and 
understand the timing behaviour between glitches \citep{alpar84a,erbil20,erbil22}.
% --- From Page 4.jpeg ---
An alternative class of models, with `star-quakes' ascribes pulsar glitches to the sudden breaking of the solid crust and the consequent decrease in moment of inertia leading to 
the spin-up observed as glitches \citep{baym71}. Since both angular momentum and the external pulsar torque on the neutron star are constant during a glitch, a pure 
star-quake would require that the fractional changes in the rotation and spin-down rates would be equal, both being due to the fractional change in moment of inertia,
$\frac{\Delta \Omega}{\Omega} = \frac{\Delta \dot{\Omega}}{\dot{\Omega}} = \frac{\Delta I}{I}$, whereas $\Delta \dot{\Omega}/\dot{\Omega}$ is much larger than 
$\Delta \Omega/\Omega$, by 3 - 6 orders of magnitude in all pulsar glitches \citep{alpar96}. In addition to this fundamental point, star-quake models fail to explain large glitches and 
certain aspects of smaller glitches \citep{giliberti20,rencoret21}. Vortex unpinning avalanche models, on the other hand, pose questions about the trigger mechanism. This leads to 
hybrid models in which the unpinning avalanche is started by crust-breaking and then amplified by superfluid vortex dynamics \citep{alpar96,erbil20,layek25}.

The size distribution of pulsar glitches as sudden steps $\Delta\Omega$ and $\Delta\dot{\Omega}$ in the rotation and spin-down rates have distinct scales, well above the timing noise 
due to the fluctuations of the continuous vortex creep process and of the external torque on the neutron star from the pulsar emission. Pulsar glitches also have a very rapid rise 
time, so far unresolved, with an upper limit of 12.6 s for one of the glitches occurred in the Vela pulsar \citep{ashton19}. Treatment of pulsar glitches therefore requires 
breaking of scale-invariance, likely imposed by scales in the distribution of pinning energies and in the scales of the underlying solid or template that provides the pinning 
centres. We present a simple outline for a model of vortex unpinning avalanches based on the breaking of scale-invariance. These considerations are relevant for the design of 
laboratory experiments and computer simulations of pulsar glitches, as well as for understanding of sudden spin-up or vorticity-current jump events in laboratory superfluids and 
condensates with pinned vortex or quantised flux line distributions. 

In Section 2, the equations of motion for vortex lines and the equations for macroscopic dynamics of a rotating superfluid are reviewed, and crust breaking is discussed. 
Section 3 presents the vortex avalanche model. The paper ends with Section 4, ``Discussion and Conclusions''.
\section*{2. SUPERFLUID ROTATION, VORTEX DYNAMICS, PINNING, DISSIPATION and CREEP}

\subsection*{2.1 Superfluid Rotation}

Superfluids rotate by sustaining quantised vortices. The macroscopic velocity field $\mathbf{v}_{\rm s} = \mathbf{\Omega}_{\rm s} \times \mathbf{r}$ of the superfluid is related to the vortex density $n(r)$ and the vorticity $\kappa$ through Stokes theorem \citep{alpar84a}:
\begin{equation}
\int_0^r (\mathbf{\nabla} \times \mathbf{v}_{\rm s}) \cdot d\mathbf{S} = \int_0^r n(r') \kappa \, 2\pi r' dr' = \oint_r \mathbf{v}_{\rm s} \cdot d\mathbf{l} = 2\pi r^2 \Omega_{\rm s}, \tag{1}
\end{equation}
or, in differential form
\begin{equation}
n(r)\kappa = 2\Omega_{\rm s} + r \frac{\partial \Omega_{\rm s}}{\partial r}. \tag{2}
\end{equation}
For rigid rotation, which is the minimum free energy state, this reduces to
\begin{equation}
n_0 \kappa = 2\Omega_{\rm s}, \tag{3}
\end{equation}
corresponding to a vortex spacing
\begin{equation}
\ell_{\rm v} \equiv \left( \frac{\kappa}{2\Omega_{\rm s}} \right)^{1/2}. \tag{4}
\end{equation}
When there are pinning forces keeping the vortex lines stuck to the normal matter, with a correspondingly higher density than the value $n_0' \kappa = 2\Omega_{\rm c} $ that would be obtained if the superfluid and normal matter-crust were co-rotating, the rotation rate of the 
superfluid will be higher than that of the normal matter-crust, $\Omega_{\rm s} > \Omega_{\rm c}$. In a pulsar, with the crust-normal matter being spun-down under the pulsar dipole radiation torque, existence of pinning forces leads to a lag 
$\omega \equiv \Omega_{\rm s} -  \Omega_{\rm c}$ between the superfluid and crust rotation rates. The vortex line velocity $ \mathbf{v}_{\rm L} =  \mathbf{v}_{\rm c} $ for pinned vortex lines, so we obtain
\begin{equation}
\mathbf{v}_{\rm s} - \mathbf{v}_{\rm L} =\mathbf{v}_{\rm s} - \mathbf{v}_{\rm c} = (\mathbf{\Omega}_{\rm s} -  \mathbf{\Omega}_{\rm c}) \times \mathbf{r} =  \mathbf{\omega} \times \mathbf{r}.  \tag{5}
\end{equation}

\subsection*{2.2 Vortex Dynamics}

The equation of motion for a vortex in a fluid, the Magnus Equation, relates the force per unit length $\mathbf{f}$ acting on the vortex line to the relative velocity $(\mathbf{v}_{\rm L} - \mathbf{v}_{\rm s})$ of the line with respect to 
the ambient superfluid flow \citep{shaham80,alpar84a,sedrakian95}:
\begin{equation}
\mathbf{f} = \rho_{\rm s} \mathbf{\kappa} \times (\mathbf{v}_{\rm s} - \mathbf{v}_{\rm L}), \tag{6}
\end{equation}
where $\rho_{\rm s}$ is the superfluid mass density, $\mathbf{\kappa}$ is the quantised vorticity with $\kappa = h/2m_{\rm n}$ for the neutron superfluid in neutron stars, $h$ being the Planck 
constant and $m_{\rm n}$ the neutron mass; and $\mathbf{v}_{\rm s}$ is the ambient superfluid velocity. In the absence of any forces acting on the vortex core, the vortex line moves with the ambient 
superfluid. With a force $\mathbf{f}$ per unit length, the relative velocity $\mathbf{v}_{\rm L} - \mathbf{v}_{\rm s}$ is in a direction perpendicular to both $\mathbf{f}$ and $\hat{\kappa}$, the direction 
of the quantised vorticity.

 The local superfluid velocity $\mathbf{v}_{\rm s}$ is determined by the distribution of all vortex lines in the system.  In the absence of any macroscopic inhomogeneities in forces acting on 
 vortex lines, a uniform distribution of vortices sustains a superfluid velocity field $\mathbf{v}_{\rm s} \cong \mathbf{\Omega}_{\rm s} \times \mathbf{r}$, approximating rigid rotation about the star's 
 rotation axis on macroscopic scales $ \gg \ell_{\rm v}$. Vortex lines pinned  to the crust-normal matter system will move with the crust; the line velocity $\mathbf{v}_{\rm L} = \mathbf{v}_{\rm c}$, the 
 velocity of the crust. The direction of the pinning force is perpendicular to both $(\mathbf{v}_{\rm c} - \mathbf{v}_{\rm s})$ and  $\mathbf{\kappa}$. As the star's outer crust and components coupled 
 tightly to the outer crust are being spun down directly by the pulsar torque, while the superfluid velocity is not decreasing because of pinning, $v_{\rm c} \le v_{\rm s}$, with 
 $\mathbf{v}_{\rm c} - \mathbf{v}_{\rm s}$ in the azimuthal direction, $\mathbf{v}_{\rm c} - \mathbf{v}_{\rm s} \parallel -\hat{\phi}$, if we take $\hat{\kappa} = \hat{\Omega}_{\rm c} = \hat{\Omega}_{\rm s}$ to be in the $\hat{z}$ 
 direction. The pinning force $\mathbf{f}_{\rm p}$ is in the $-\hat{r}$ direction, towards the rotation axis, in this case of uniform vortex distribution. We are using cylindrical coordinates 
 $(r, \phi, z)$ with $\hat{z} = \hat{\Omega}$, the rotation axis. The maximum, critical velocity difference between a vortex line and the ambient superfluid, $|\mathbf{v}_{\rm cr}| 
 \equiv  |\mathbf{v}_{\rm s} - \mathbf{v}_{\rm L}|_{\rm max}$ corresponds to the maximum available pinning force:  
 \begin{equation}
 {f}_{\rm p,max}  \equiv  \rho_{\rm s} \kappa v_{\rm cr} \tag{7},
\end{equation}
which defines a critical lag 
\begin{equation}
 \omega_{\rm cr}\; r  \equiv v_{\rm cr}. \tag{8}
\end{equation}

When a vortex line is unpinned, it will start motion with $\mathbf{v}_{\rm L} \cong \mathbf{v}_{\rm c} = \mathbf{\Omega}_{\rm c} \times \mathbf{r}$. Interaction of the vortex line core with electrons and phonons 
will cause an azimuthal drag force. In response, the vortex line will acquire a small radially outward velocity $v_{\rm L,r}$ away from the rotation axis. The small angle $\zeta \equiv 
\frac{v_{\rm L,r}}{v_{\rm L,\phi}}$ describes this outward motion of unpinned vortex lines.

\subsection*{2.3 Vortex Creep}

Pinned vortices can get unpinned and re-pinned by thermal activation against pinning potential wells and barriers \citep{alpar84a,alpar89}. As the crust and normal matter are being spun down by the pulsar 
torque, vortex pinning leads to a lag $\omega \equiv \Omega_{\rm s} - \Omega_{\rm c}$. There is a bias favouring radially outward vortex motions in the $\hat{r}$ direction, which reduce the lag, 
compared to motion in the $-\hat{r}$ direction. This leads to a vortex current in the $\hat{r}$ direction, achieved by the dissipative azimuthal drag forces on unpinned vortices. 
There is a steady-state of creep dynamics, in which the radial vortex motion allows the superfluid to spin-down at the same rate as the normal matter-crust, sustained by a steady-state value $\omega_{\infty}$ of the lag. 

Vortex creep has a non-linear exponential dependence of rates on $\omega/T$ through Boltzmann factors with $T$ being the internal temperature of the neutron star. The steady-state value of the velocity difference is 
$v_{\infty} = v_{\rm cr} [ 1 -  (k_{\rm B} T/E_{\rm p}) \ln] $, where $k_{\rm B}$ is the Boltzmann constant, $R$ is the neutron star radius, $\ln \equiv \ln [2 v_0 \Omega / (R |\dot{ \Omega}|)] \cong 30$ and $v_0$ is a typical speed of a vortex line between pinning sites \citep{alpar89,erbil16}. 
%The steady-state lag and critical lag for unpinning are related by $\omega_{\infty} = \omega_{\rm cr} [ 1 - (k_{\rm B} T/E_{\rm p}) \ln ]$.

\section*{3. THE AVALANCHE MODEL}

\subsection*{3.1 Vortex Traps and the Start of the Avalanche} 

The distribution of pinned vortices is re-arranged, under stresses arising from spatial variations of the pinning force, into domains of size $D$ determined by the 
critical strain angle $\theta_{\rm cr}$ of the crystalline or amorphous solid \citep{akbal15,erbil19}:
\begin{equation}
D \cong \delta R \theta_{\rm cr}, \tag{9}
\end{equation}
where $\delta R \sim 10^5$ cm is the solid crust thickness. The critical strain angle in the neutron crust solid is much larger than in terrestrial solids. This is because of the 
lack of screening: The electrons in the neutron star crust are relativistic, with negligible Coulomb energies $Ze^2/a$, where $Z$ is the charge on ions, $e$ the electron charge and  
$a$ the inter-ion spacing, compared to the relativistic Fermi energy $\hbar k_{\rm F} c$, where $\hbar$ and $c$ are the reduced Planck constant and the speed of light, and 
$k_{\rm F}$ is the electrons' Fermi wave-number. The dimensionless critical strain angle $\theta_{\rm cr}$ is expected to be of the order \citep{alpar89a}
\begin{equation}
\theta_{\rm cr} \cong \alpha \frac{Z^{4/3}}{(3 \pi^2)^{1/3}} \sim O(10^{-1}), \tag{10}
\end{equation}
where $\alpha$ is the fine structure constant and $Z$ is the number of protons in the nuclei making up the solid, which is in agreement with the results of recent 
numerical simulations \citep{baiko18,morales25}. 

The vortices clustered, with an areal density $n_0 + \delta n$, into a domain, which we call a vortex ``trap'', of size $D$ would have been distributed with density $n_0$ in a domain 
of size $D' > D$ prior to the formation of the trap, leading to 
\begin{equation}
\frac {{D' }^2} {{D}^2} = \frac {n_0 + \delta n} {n_0}. \tag{11}
\end{equation}

The circulation around the trap region can at most lead to a critical velocity difference $\mathbf{v}_{\rm cr}$ between 
the velocity $\mathbf{v}_{\rm L} = \mathbf{v}_{\rm c}$ of pinned vortices moving with the velocity $\mathbf{v}_{\rm c}$ of the crust, and the local superfluid velocity 
$\mathbf{v}_{\rm s}$. The critical speed corresponding to the maximum (available) pinning force is given by equation (7), in accordance with the Magnus force, equation (6). 
The increased vortex density $(n_0 + \delta n)$ in a trap region sustains an extra, local superfluid flow, with magnitude $\delta v_{\rm s} = \bar{\Omega} r'$, mimicking rigid 
rotation around the centre of the trap, with circulation $2 \bar{\Omega}=(n_0 + \delta n) \kappa$,  $r'$ being the distance from the centre of the trap. The background superfluid 
velocity due to the large-scale average uniform distribution of vortices is $v_{\rm s} = \Omega_{\rm s} r$.  For pinning in the inner crust of the neutron star, the distance $r$ from 
the rotation axis is close to the stellar radius, $r \cong R_{\*}$. The magnitude of the difference between the superfluid velocity and the velocity 
$\mathbf{v}_L = \mathbf{v}_{\rm c}$ of the vortex lines pinned to and co-rotating with the crust-normal matter reaches the critical value $v_{\rm cr}$ at the boundaries of the vortex 
trap. This difference is
\begin{equation}
\mathbf{v}_{\rm s} - \mathbf{v}_{\rm c} = \mathbf{\Omega}_{\rm s} \times \mathbf{r}_{\rm trap}+ v_{\rm cr} \frac{r' }{D/2} \hat{\theta} -   \mathbf{\Omega}_{\rm c} \times \mathbf{r},  \tag{12}
\end{equation}
where $\hat{\theta}$ is the direction of circulation on the boundary of the trap, and $\mathbf{r}_{\rm trap}$ is the position of the centre of the  trap, so that
\begin{equation}
\bar{\Omega} = \frac{v_{\rm cr} }{D/2}. \tag{13}
\end{equation}
For a circular trap with radius $D/2$,
\begin{equation}
(n_0 + \delta n) \kappa = 2 \bar{\Omega}= 4  \frac{v_{\rm cr}}{D} ,  \tag{14}
\end{equation}
by Stokes' theorem, since the circulations of $\mathbf{\Omega}_{\rm s} \times \mathbf{r}$ and $\mathbf{\Omega}_{\rm c} \times \mathbf{r}$ around the local trap both vanish. It 
follows  from equations (3), (11) and (14) that $\bar{\Omega}/\Omega = (D'/D)^2$. 

Thus, the physics of the crust solid and the maximum pinning forces available set the scales $D$ and $D'$ of the traps and surrounding vortex-free regions, according to equations 
(9)-(14). Within each trap vortices creep outward from the centre of the trap, in the $\hat{r'}$ direction. In steady-state, the creep within the trap is sustained by a superfluid 
flow
\begin{equation}
\mathbf{v}_{\rm s,\infty} - \mathbf{v}_{\rm c} = \mathbf{\Omega}_{\rm s} \times \mathbf{r} + v_{\infty} \frac{r'}{D/2} \hat{\theta} -   \mathbf{\Omega}_{\rm c} \times \mathbf{r}.  \tag{15}
\end{equation}
Vortices from a particular trap will unpin when the velocity difference between the superfluid and the pinned vortex lines reaches $v_{\rm cr}$ on the outer boundary of the trap. 
As noted above,  the superfluid velocity $v_{\infty}$ in steady-state vortex creep at the boundaries of the trap is slightly less than the critical value $v_{\rm cr}$,
\begin{equation}
v_{\infty} = v_{\rm cr} [ 1 -  (k_{\rm B} T/E_{\rm p}) \ln [2 v_0 \Omega / (R |\dot{ \Omega}|)]. \tag{16}
\end{equation}
For typical neutron star interior temperatures $k_{\rm B} T \sim 10$ keV \citep{,potekhin20} and typical pinning energies $E_{\rm p} \sim 1$ MeV \citep{ruderman76,seveso16} while 
$\ln [2 v_0 \Omega / (R |\dot{ \Omega}|)] \cong 30$ \citep{alpar89}. Thus, the steady-state velocity $v_{\infty}$ is only about 30\% less than $v_{\rm cr}$. Once this difference is 
overcome by some fluctuation in a particular vortex trap, the unpinned vortices start an avalanche, triggering vortex unpinning from other traps they encounter and thus producing 
the pulsar glitch \citep{cheng88}.     
 
Now, let us turn our attention to the motion of free (unpinned) vortices. The relative velocity between free vortex lines and normal matter, in the azimuthal direction with respect to 
the rotation axis, is faced by dissipative drag forces, due to interactions of vortex line cores, which contain normal matter, with the electrons and phonons co-rotating with the 
crust \citep{bildsten89}. Energy dissipation by excitation of Kelvin modes (kelvons) along vortex lines also contribute to the drag forces \citep{epstein92,jones92}. The azimuthal drag forces 
result in a small radial motion of the unpinned lines, described by a ``dissipation angle''
\begin{equation}
\zeta \equiv \frac{V_{\rm L,r}}{V_{{\rm L},\phi}} \cong \frac{V_{\rm L,r}}{\Omega R} = \frac{dr}{r d\phi} \cong \frac{\Delta R}{2 \pi R}, \tag{17}
\end{equation}
where $\Delta R$ denotes the radial distance moved by free vortex lines during one azimuthal turn, of length $2 \pi R$; here $R$ is the distance of the pinned superfluid from the 
rotation axis, approximately equal to the neutron star's radius. When vortices in one trap get unpinned, they will move azimuthally, slightly spacing out radially also, tracing 
tightly wound spiral trajectories \citep{sedrakian95,haskell16}. They will encounter other vortex traps, increasing the local superfluid velocity in these encounters and triggering 
unpinning from the vortex traps they encounter, leading to an avalanche \citep{alpar96,erbil20}. Individual unpinned vortices will eventually settle to basically azimuthal motion 
when they do not encounter traps, so not all unpinned vortices move all the way out of the superfluid. The avalanche is rather like a relay race, in which individual vortices may 
participate for a while, moving rapidly through vortex-free regions, encountering and triggering release of vortices from vortex traps on their way and eventually re-pinning in a 
vortex trap, before travelling the full radial extent $\delta R$ of the pinned superfluid. The avalanche leads to a reduction $\delta \Omega$ in the superfluid rotation rate, which 
is related to the radial distance $\Delta R$ covered in azimuthal turn by
\begin{equation}
\frac{\delta \Omega}{\Omega} =  2 \frac{\Delta R}{R} . \tag{18}
\end{equation}

Assuming that by the time $ t_{\rm g}$ to the next glitch vortex creep has healed to replenish this reduction in the superfluid rotation rate so that conditions are ripe for the 
next unpinning avalanche-glitch, we obtain \citep{alpar84a,alpar89}
\begin{equation}
\delta \Omega =  |\dot\Omega | t_{\rm g}. \tag{19}
\end{equation}

From equations (17)-(19) we obtain
\begin{equation}
\zeta =\frac{\delta \Omega}{4 \pi\Omega} =  \frac{1}{8 \pi} \frac{t_{\rm g}}{\tau_{\rm sd}},  \tag{20}
\end{equation}
where $\tau_{\rm sd} \equiv \Omega/(2|\dot{\Omega}|)$ is the pulsar's spin-down timescale.

\subsection*{3.2 The Trap Landscape and the Development of the Avalanche}

\begin{figure}
    \centering
    \includegraphics[width=1\linewidth]{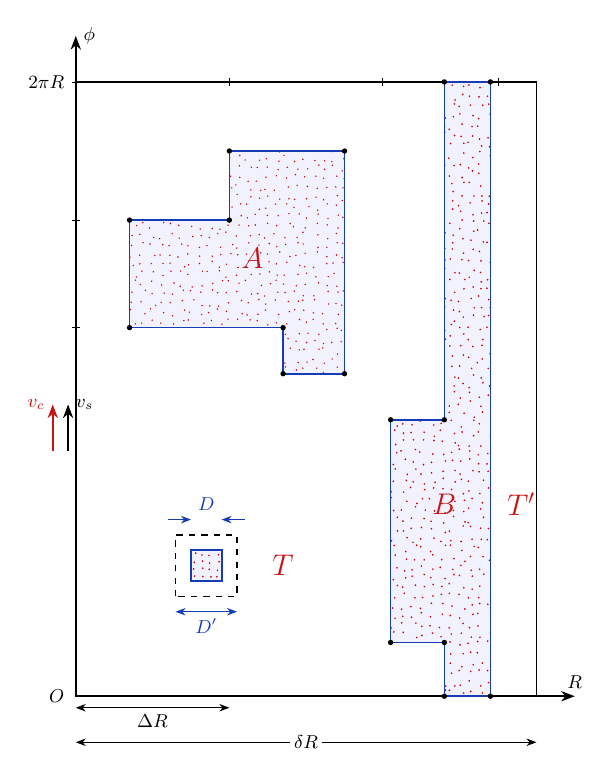}
    \caption{Schematic description of the crust superfluid with a partially developed landscape of vortex traps. The model is axially symmetric with respect to the star's rotation 
    axis. The crust superfluid, a ring of circumference $2\pi R$ and width $\delta R$ in planes perpendicular to the rotation axis, is depicted as a strip of length $2\pi R$ and 
    width $\delta R$. Rectangular geometry is chosen for simplicity. Each vortex trap, a square of side $D$, at the centre of a square "tile" of side $D' > D$, is surrounded by vortex-free space. Dots represent the pinned vortices, engaged in vortex creep. Two examples of extended vortex pinning regions are shown. Region $A$ is disjoint in the azimuthal 
    direction. Region $B$ is connected azimuthally, providing a barrier separating the tiled regions $T$ and $T'$. Only one representative tile is shown to avoid clustering the figure. $\Delta R \ll \delta R$ is the outward radial distance covered by each free vortex, per azimuthal turn $2 \pi R$, due to drag forces (mutual friction).}
    \label{fig:ava}
\end{figure}

Fig. \ref{fig:ava} shows the pinned crust superfluid as a strip of length $2 \pi R$ and width $\delta R$. This is tiled into squares of size $D'$, of which a fraction $\eta$ have 
undergone trap formation. Unpinned vortices have the radial reach $\Delta R$ per azimuthal turn $2 \pi R$. Vortex creep proceeds within traps and in any extended regions where traps 
have not been formed; while unpinned vortices will move rapidly, with the local superfluid velocity, through the vortex-free regions between traps. The traps are taken to be squares 
of side $D$ with the superfluid circulation shown. [It so happens that equation (14) holds also for circular traps - for other trap geometries a factor $\sim O(1)$ would be needed.] 
The cross section for a trap to interact with the streaming unpinned vortices is the geometrical cross section $\sigma = D$, as a streaming vortex would need to come extremely 
close, within a distance $\delta \ll D$ to the trap for its contribution $\delta v = \kappa / (2 \pi \delta)$ to the superfluid velocity to be a significant fraction of $v_{\rm cr}$. 
Using the density of traps $n_{\rm trap} = 1/ D'^{2}$ we obtain the mean free path 
\begin{equation}
l_{\rm mfp}= \frac{D'^2}{D}. \tag{21}
\end{equation}
For unpinned vortices, the number of encounters with traps in one azimuthal turn is
\begin{equation}
N_{\rm enc} =  \frac{ \eta 2\pi R }{l_{\rm mfp}} =  \eta 2\pi R \frac{D}{D'^2}. \tag{22}
\end{equation}
The average number of vortices unpinned from each encountered trap is a small fraction of the number of vortices $N_{\rm trap}= 4 D v_{\rm cr}/\kappa $ pinned in the trap:
\begin{equation}
N_{\rm unpin} \ll  N_{\rm trap} = 4 D v_{\rm cr}/\kappa.   \tag{23}
%\frac{N_{\rm Ava}}{N_{\rm enc}} =   \frac{\zeta}{\eta}  \frac{ 8 \pi R v_{\rm cr}}{\kappa}, 
\end{equation}
leading to a bound on the total number $N_{\rm Ava}$ of vortices participating in the avalanche: 
\begin{equation}
N_{\rm Ava}  < N_{\rm enc} N_{\rm unpin} \frac{\delta R}{\Delta R} \ll N_{\rm enc} N_{\rm trap} \frac{\delta R}{\Delta R}  = \frac{\eta}{\zeta} \left(\frac{D}{D'}\right)^2 \frac{4 v_{\rm cr}\delta R}{\kappa}, \tag{24}
\end{equation}
%\frac{\zeta}{\eta}    \frac{ 2 \pi R}{D} < 1  ( \ll  1 ) 
through equations (17), (22) and (23). Angular momentum transfer from the pinned superfluid, of moment of inertia $I_{\rm s} $, to the crust-normal matter during the avalanche leads to the observed sudden increase 
$\Delta \Omega$ in the rotation rate of the crust-normal matter system, with moment of inertia $I_{\rm c} \cong I$, the total moment of inertia of the star \citep{alpar84b}. The pinned superfluid 
is made up of vortex creep regions, including the creep regions inside vortex traps, of total moment of inertia $I_{\rm A}$, and vortex-free regions between traps, of total moment of  
inertia $I_{\rm B}$ in the notation of vortex creep literature \citep{alpar96,alpar06,erbil20,erbil22}. We obtain
\begin{equation}
\beta \equiv \frac{I_{\rm B}}{I_{\rm A}} = \frac{\eta[1 - (D/D')^2]}{1 -  \eta[1 - (D/D')^2]}. \tag{25}
\end{equation}
As expected, $\beta$ increases with $\eta$, the fraction of superfluid that is tiled into vortex traps surrounded with vortex-free regions, and with the ratio $[1 - (D/D')^2]$ of the 
vortex-free area to the tile area in each tile where vortices have been clustered into a trap. 

Note that the fast, ballistic motion of free vortices in the azimuthal direction means that the progress of the avalanche is not a percolation process in the two-dimensional 
landscape of vortex traps and extended vortex pinning-creep regions; the avalanche propagates essentially in the radial direction. There is no percolation in one-dimensional transport. 
This means that azimuthally connected extended pinning-creep regions, like the region $B$ shown in Fig. \ref{fig:ava} pose a topological barrier to the growth and reach of the avalanche and size of the resulting glitch: an avalanche originating in either side of region $B$, in the zones $T$ and $T'$ will be confined to its zone of origin.

The avalanche reduces the superfluid rotation rate by the average value $\bar{\delta\Omega} = \delta\Omega/2$ in vortex creep regions and disrupts the creep process, leading to the 
observed step increase $\Delta\dot{\Omega}$ in the spin-down rate, which gives the moment of inertia of the vortex creep regions \citep{alpar06}
\begin{equation}
\frac{\Delta \dot{\Omega}}{\dot{\Omega}} = \frac{I_{\rm A}}{I_{\rm c}}. \tag{26}
\end{equation}
Angular momentum conservation thus yields
\begin{equation}
I_{\rm A}\frac{\delta \Omega}{2} + I_{\rm B} \delta \Omega = I_{\rm c} \Delta \Omega,  \tag{27}
\end{equation}
using equation (19). From equations (26) and (27) one can express the glitch magnitude as
\begin{equation}
    \frac{\Delta\Omega}{\Omega}=\left(\beta+\frac{1}{2}\right)\frac{\Delta\dot\Omega}{\dot\Omega}\frac{t_{\rm g}}{2\tau_{\rm sd}}, \tag{28}
\end{equation}
in terms of $\beta\equiv I_{\rm B}/I_{\rm A}$, inter-glitch time $t_{\rm g}=\delta\Omega/|\dot{\Omega}|$ and spin-down timescale $\tau_{\rm sd}=\Omega/(2|\dot{\Omega}|)$. We can now extract the value of $\beta$ from the observed glitch steps $\Delta\Omega$ and $\Delta \dot{\Omega}$, and the time $ t_{\rm g} $ to the next glitch. 

For the Vela pulsar, the parameter $\beta$ has been determined for individual glitches. To order of magnitude, the glitch jumps are $\Delta \Omega \sim 10^{-4}$\; \mbox{rad s$^{-1}$}
and $ |\Delta\dot{\Omega}| \sim 10^{-13}$\; \mbox{rad s$^{-2}$}. The average interval between glitches is 1000 days, $ t_{\rm g} \sim 9 \times 10^7$\; s.  Detailed timing fits covering 18 Vela pulsar glitches using the vortex-creep model have revealed that $\beta$ varies within the range of 1.18–4.96 \citep{akbal17,erbil22}. This yields an average $\langle\beta\rangle = 2.2$ for these glitches with $\langle t_{\rm g}\rangle = 1060$ days. Equation (25) yields $ \eta \; [ 1\; -\frac {{D }^2} {{D'}^2} ] \simeq 0.7$, which means that $\eta \lesssim 1 $, i.e. trap formation is almost complete in the Vela pulsar.

For the Crab pulsar $\beta \sim 0.25 $ was obtained for the largest glitch \citep{erbil19}, and similar values were obtained for other glitches \citep{alpar96}, leading to 
$\eta \; [ 1\; -\frac {{D }^2} {{D'}^2} ] \sim \beta \sim 0.25$.

 For PSR J0537--6910, a range of 1.46-32.77 was obtained for $\beta$ with an average of $\langle\beta\rangle = 13.2$ from timing fits covering 38 glitches using the vortex-creep model \citep{erbil26}. The average value $\langle\beta\rangle = 13.2$ gives, by equation (25), $ \eta \; [ 1\; -\frac {{D }^2} {{D'}^2} ] \simeq 0.93$. 
 
These results are in agreement with the 
notion that trap formation is an evolutionary process which is incomplete in the young Crab pulsar, and advances as the pulsar ages, nearing completion by the characteristic age of
PSR J0537--6910, $\cong 4900$ years, and in Vela ($\tau_{\rm sd}=11.4$ kyr) and older pulsars. 

\subsection*{3.3 Determination of the Avalanche Parameters from Input Physics and Observed Glitch Parameters}

Starting with the choice of $\delta R$, the radial extent (or thickness) of the pinned superfluid in the neutron star crust, and using as physical input the critical strain angle for the Coulombic neutron star solid with relativistic electrons, we obtain the typical size $D$ 
of a solid plate that acts as a vortex trap from equations (9) and (10). Using the maximal value of the pinning force per unit vortex length, $f_{\rm p}$ that imposes the extra vortex density in the trap, and the corresponding critical velocity $v_{\rm cr}$ at the boundaries of the trap, 
the contrast factor $ {D' }^2 / {D}^2 $, i.e. the increase in vortex density in the traps is obtained from equations (3), (9), (11) and (14):
\begin{equation}
\frac {{D' }^2} {{D}^2} = \frac {n_0 + \delta n} {n_0} = \frac{ 2 v_{\rm cr}}{D\; \Omega} = \frac{ 2 v_{\rm cr}}{\delta R\; \theta_{\rm cr}\; \Omega}. \tag{29}
\end{equation} 
The fraction $\eta$ of the pinned superfluid regions where traps are present and the density contrast ${D' }^2/{D}^2$ are related to the ratio $\beta$ of the moments of inertia of the vortex-free regions and regions carrying pinned vortices and vortex creep, through equations (25) and (28)
\begin{equation}
\eta \; [ 1\; -\frac {{D }^2} {{D'}^2} ] = \frac {\beta} {\beta \; +\; 1} = \frac{ (\frac{\Delta \Omega}{ |\Delta\dot{\Omega}| t_{\rm g }} \; - \frac{1}{2} )}{(\frac{\Delta \Omega}{ |\Delta\dot{\Omega}| t_{\rm g } } \; + \frac{1}{2} )}  \tag{30},
\end{equation}
completing the avalanche solution corresponding to the physical parameters and observed glitch parameters of a particular neutron star (or a pinned vortex or flux system).

\subsection*{3.4 Application to Pulsar Glitches}

The maximum pinning force per unit length of a vortex line in the neutron star pinned superfluid is
\begin{equation}
f_{\rm p} = \frac{E_{\rm p}}{ \xi \; b } \sim \frac{E_{\rm p}\;(\mbox{MeV})}{\xi (10 \;\mbox{fm})\; b ( 100 \; \mbox{fm})} \; \times 10^{17} \; \mbox{dyne cm$^{-1}$}.  \tag{31}
\end{equation}
Here $E_{\rm p}$ is the pinning energy at each site where the vortex line is pinned to a nucleus in the crust \citep{alpar77,epstein88,seveso16}, $\xi \sim 10$ fm is the radius of 
the vortex line (the coherence length of the superfluid), and $b$ is the distance between successive pinning sites along the vortex line, estimated to be $b\sim 100$ fm in the weak 
pinning regime which is likely to prevail in most parts of the inner crust solid-pinned superfluid \citep{shaham80,anderson82}. Weak pinning means the pinning forces are not strong 
enough to dislodge nuclei from their equilibrium sites, so that vortices pin at geometrically random sites, at average spacing $b \cong a^3 / (\pi \xi^2)$ along each vortex line, $a$ 
being the average spacing between nuclei. The critical velocity corresponding to the maximum pinning force is 
\begin{equation}
v_{\rm cr} = \frac{f_{\rm p,max}}{ \rho_{\rm s} \; \kappa } \sim \frac{f_{\rm p,max,17}}{ \rho_{\rm s,13}} \; \times 10^7\; \mbox{cm s$^{-1}$}, \tag{32}
\end{equation}
where $f_{\rm p,max,17}$ is the pinning force per unit length of the vortex line in units of $ 10^{17} $\; \mbox{dyne cm$^{-1}$}, and $\rho_{s,13}$ is the superfluid density in 
units of $10^{13}$\;\mbox{gr cm$^{-3}$}. From equations (10), (29) and (32) we obtain 
\begin{equation}
\frac {{D' }^2} {{D}^2} \sim 30 \frac{ v_{\rm cr,7}}{\delta R_5 \Omega_{70}\;}, \tag{33}
\end{equation}
with the angular speed is scaled to $\Omega\cong70$\;\mbox{rad s$^{-1}$}, the rotation rate of the Vela pulsar. 

For the Coulombic neutron star crust solid the shear modulus $\mu$ is given by \citep{strohmayer91} 
\begin{equation}
\mu \cong \frac {Z^2 \; e^2} {a^4} \sim 10^{29} - 10^{30}\; \mbox{erg cm$^{-3}$}, \tag{34}
\end{equation}
where $Z$ is the atomic number (nuclear charge).
The critical stress $\sigma_{\rm cr}$ provided by maximum pinning force at one pinning site per area $b^2$ is consistent with the shear modulus of the Coulomb solid \citep{ruderman76}:
\begin{equation}
\sigma_{\rm cr } \equiv \mu \theta_{\rm cr} \sim   \frac {f_{\rm p,max}} {b} \sim 10^{28} - 10^{29}\; \mbox{erg cm$^{-3}$}. \tag{35}
\end{equation}

According to equation (20), the radial scattering angle $\zeta$ is expected to be of order $10^{-4} - 10^{-5}$ for pulsars with inter-glitch intervals $t_{\rm g} \sim 1 $ yr and spin-down age 
$\tau_{\rm sd} \sim 10^3 -10^4 \;$ yr.

The avalanche will take place in a time
\begin{equation}
{\Delta t}_{\rm Ava} = \frac{ \delta R}{v_{\rm r}} = \frac{ \delta R}{\zeta \Omega R} \tag{36},
\end{equation}
 the glitch rise time. From the best resolved glitch timing data for the Vela pulsar we have the limit
 \begin{align}
{\Delta t}_{Ava} &<  12.6 \; \mbox{s}, \nonumber \\
\frac{\delta R}{\Omega R \times 12.6 \;\mbox{s}} &< \zeta , \nonumber \\
10^{-5}\frac{{\delta R}_4}{R_6}   &< \zeta,   \tag{37}
\end{align}
using $\Omega \cong 70 \; \mbox{rad s$^{-1}$} $ for the Vela pulsar.

Radial scattering due to the interaction of vortex lines with phonons \citep{jones90,haskell12} and due to Kelvin wave excitations (kelvons) on the vortex lines 
\citep{epstein92,jones92,haskell12, graber18} have been calculated. Considering both mechanisms, with pinning energies $E_{\rm p} \sim 1 $ MeV, \citet{haskell12} find that in the 
inner crust phonon and kelvon drag make comparable contributions to mutual friction, leading to $\zeta \sim 10^{-5}$ at densities $\rho \cong 4 \times 10^{13}$ \mbox{g cm$^{-3}$}, while at the 
highest crust densities $\rho \cong 8 \times 10^{13}$ \mbox{g cm$^{-3}$} kelvon drag dominates, giving $\zeta \sim 10^{-3}$. 

\section*{4. Discussion and Conclusions}

We have developed a simple model for the development of pulsar glitches as avalanches of vortex unpinning in the inner crust superfluid of neutron stars. Vortex unpinning is triggered 
in a vortex trap, which holds a high density of pinned vortices. Each trap is surrounded by vortex-free regions with high local superfluid velocity circulating around the trap. 
Vortices unpinned from any particular trap will move rapidly through the vortex-free regions, essentially in the azimuthal direction around the neutron star's rotation axis, while 
also moving radially outward, at a much slower rate compared to their azimuthal motion with the superfluid flow. The radial motion is caused by the dissipative interactions of vortex 
lines with ambient normal matter and energy transfer to vortex line excitations, azimuthal drag forces leading to radial vortex motion in accordance with the Magnus equation of 
motion for vortices. 

The scales of vortex traps and surrounding vortex-free regions are provided by two physical parameters: 

(i) The breaking strength of the crust solid: the neutron star crust solid is 
an unscreened Coulomb lattice due to the relativistic kinetic energies of the electrons. The dimensionless parameter that defines the critical strain of the solid, set by the fine 
structure constant and the nuclear charge $Z$, is likely to be of order $10^{-1}$, as demonstrated by molecular dynamics simulations. 

(ii) The critical (maximum) difference between superfluid velocity and the velocity of vortices pinned to the normal matter solid that can be sustained by the maximum 
available pinning force per unit length. Vortices unpin when this critical velocity difference is reached. Although the pinning energies and the distances between successive 
pinnings along a vortex line vary throughout the crust superfluid, typical pinning energies are of order $E_{\rm p} \sim 1 $ Mev, and the distance $b$ between successive pinning 
sites is expected to be of order $b \sim 100$ fm. 

These physical scales determine the size of traps and the extra vortex density they sustain. The traps are formed by crust breaking events, which themselves are triggered by 
stresses arising from pinning forces. When a young neutron star cools down through the superfluid phase transition, within a few hundred years of its formation \citep{ho15}, it 
will start with a spatially connected pinning landscape and a uniform vortex density \citep{sauls26}. As more vortex traps are formed with time, the pinned superfluid 
becomes ``tiled''  into vortex traps and surrounding vortex-free regions to a progressively greater extent \citep{cheng88}. The sizes of vortex avalanches, the extent of the tiled 
pinning landscape through which the avalanche moves and therefore the glitch sizes will grow with a pulsar's age, eventually reaching the full extent of the pinned crust superfluid, 
corresponding to a maximum glitch size \citep{alpar06,erbil20}. Comparison of the observed glitch parameters of the Crab and Vela pulsars and PSR J0537--6910 shows that our avalanche 
model is in qualitative agreement with this evolutionary picture. 

The vortex avalanche moves in the radially outward direction by virtue of the dissipative forces acting on the free vortex lines, described by mutual friction 
parameters \citep{haskell15}. With the hypothesis that repeated glitches occur when similar conditions are reached, we make an estimate of the radial scattering parameter which is in 
agreement with current calculations of the mutual friction parameters. In addition, a lower bound on the radial scattering parameter obtained from the best resolved limit on the 
glitch rise time in the Vela pulsar is also in agreement with the calculated mutual friction parameters. 

A vortex avalanche can proceed in a fast timescale only if there are pathways for fast vortex motion carried along with the superfluid flow. This is not possible within vortex 
pinning regions where gradual radially outward vortex motion is realized diffusively and continuously in the vortex creep process, on the spin-down timescale dictated by the 
external torque. The scale-invariance in an extended homogeneous pinning landscape is broken if the scale-invariance in the spatial distribution of pinned vorticity is broken by the 
formation of local vortex traps surrounded by vortex-free regions which can connect to form pathways for fast vortex motion. The breaking of spatial space-invariance also leads to 
a qualitative understanding of the bimodality in the observed distribution of pulsar glitch magnitudes. 

We conclude by emphasizing that the design of experiments and simulations on vortex systems to emulate pulsar glitches should include spatially inhomogeneous vortex pinning 
landscapes. Breaking the spatial scale-invariance and monitoring the results will provide more realistic pulsar glitch simulations. This approach will also lead to better 
understanding of collective behaviour in pinned vortex systems in general.

\section*{acknowledgements}

This article benefited from meetings and discussions within COST Action SCALES, CA24139, supported by COST (European Cooperation in Science and Technology). EG is supported by the Doctor Foundation of Qingdao Binhai University (No. BJZA2025025).

\section*{Data Availability}

No new data were generated in support of this theoretical study.

%%%%%%%%%%%%%%%%%%%%%%%%%%%%%%%%%%%%%%%%%%%%%%%%%%%

\label{lastpage}

\end{document}